\documentclass[aps,8pt,twocolumn,tightenlines,floatfix]{revtex4}
\usepackage[dvips]{graphicx}
\usepackage[english]{babel}
\usepackage{amsmath}
\usepackage{amssymb}
\usepackage{slashed}
\usepackage{color}
\usepackage{times}
\usepackage{verbatim}
\usepackage[english]{babel}

\newcommand{\uk}{u_{\mathbf{k}}}
\newcommand{\vk}{v_{\mathbf{k}}}

\def\beq{\begin{eqnarray}} \def\eeq{\end{eqnarray}}

\begin{document}

\title{Theory of THz Conductivity 
in the Pseudogap Phase of the Cuprates:
A Pre-Formed Pair Perspective.~~~I}

\author{Dan Wulin}
\author{Vivek Mishra}
\author{K. Levin}
\affiliation{James Franck Institute and Department of Physics,
University of Chicago, Chicago, Illinois 60637, USA}

\begin{abstract}
In this paper we deduce transport properties in the presence of a
pseudogap associated with precursor superconductivity. 
Our theoretical analysis is based on the widely adopted
self energy expression reflecting this normal state gap, which has
appeared in interpretations of
photoemission and in other experiments. Thus, it should
be generally applicable. Here we address THz conductivity
$\sigma (\omega) = \sigma_1(\omega) + i \sigma_2(\omega)$
measurements in the underdoped high temperature superconductors
and arrive at reasonable agreement between theory and recent experiment
for both $\sigma_1$ and $\sigma_2$ above and below $T_c$.
\end{abstract}
\maketitle

\section{Introduction}

One of the biggest challenges in understanding the high temperature
superconductors revolves around the origin of the ubiquitous pseudogap.
Because this normal state gap has $d$-wave like features compatible with the 
superconducting
order parameter, this
suggests that the pseudogap
is related to some form of
``precursor pairing".
On the other hand, there are many reports \cite{Taillefer4,Hinkov} suggesting
that the pseudogap onset temperature is
associated with a broken symmetry and, thus, another order parameter.
It is widely believed that because the pseudogap has clear
signatures in generalized transport,
these measurements \cite{Corson1999,Ong2} 
may help with the centrally important question of
distinguishing the two scenarios.
In this paper we analyze
recent experimental observations
which have suggested that
precursor pairing scenarios \cite{Bilbro,Bilbro2} may
be problematic. By
default, these observations may imply
that the pseudogap must involve
another (yet unspecified) order parameter.

Our work is based on a preformed pair scenario \cite{CSTL05}
which has been previously applied to transport
\cite{Chen2,Arcstransport,NJOP,Ourviscosity,OurBraggPRL}
within a slightly different, but equivalent framework.
Importantly this preformed pair scheme is associated
with the widely used
\cite{CSTL05,FermiArcs,NormanTransport,SenthilLee,Chubukov2}
approximate self energy, which we derived
even earlier within our microscopic formalism \cite{Maly1,Maly2}
\begin{eqnarray}
\Sigma_{pg,K}=-i\gamma+\frac{\Delta_{pg,\textbf{k}}^2}{i\omega_n+\xi_{\textbf{k}}+i\gamma}.
\label{eq:4a}
\end{eqnarray}
where $K=(\textbf{k},i\omega_n)$ and $i\omega_n$ is a fermionic Matsubara frequency.
Here
$\gamma$ represents a damping, which
we will interpret here as related to
the inter-conversion of pairs and fermions.
We next show how this self energy leads very naturally to
an expression for the complex conductivity.

\section{Transport Theory in the Presence of a Preformed Pair-Based-
Pseudogap} \label{sec:3
s}

The complex conductivity
can be written in terms of the paramagnetic current-current correlation function
$\tensor{P}(Q)$ to which one adds the diamagnetic contribution
$\tensor{n}/m$

 \begin{eqnarray}
\sigma(\omega)=-\displaystyle{\lim_{\textbf{q}\rightarrow0}}\frac{P_{xx}(\textbf{q},\omega)+(n/m)_{xx}}{i\omega}\label{eq:conddef}
\end{eqnarray}
where the $xx$ subscript denotes the diagonal tensor component along the $x$ direction. We consider in the transverse gauge the linear response of
the electromagnetic current ${\bf {J}} = - \overleftrightarrow{K} \bf{A}$,
to a small vector potential 
$\textbf{A}$ with
$\tensor{K}(Q) = \tensor{P}(Q)+\tensor{n}/m$, where the paramagnetic
contribution, given by $\tensor{P}(Q)$, is associated with the normal current resulting from
fermionic and bosonic excitations. The vector $Q$ is defined $Q=(\textbf{q},i\Omega_m)$ where $i\Omega_m$ is a bosonic Matsubara frequency.

\subsection{Weak Dissipation Limit}
For simplicity, we begin in the weak dissipation limit where the parameter
$\gamma$ in Eq.~(\ref{eq:4a}) is small. 
We define $G_{0,K}^{-1}=(i\omega_n-\xi_{\bf k})^{-1
}$ as the bare Green's function, and
show
how this standard self energy expression in the pseudogap state,
$$\Sigma_{pg,K} \approx -\Delta_{pg,\bf k}^2G_{0,-K} = \frac{\Delta_{pg,\bf k}^2}{i\omega_n+\xi_{\bf k}}$$
leads to consistent expressions for the
current-current correlation functions, which were earlier presented
using a more microscopic formulation \cite{Kosztin2,ourreview,Chen2}. 

We derive an expression for $P(Q)$
by 
turning first to the diamagnetic current. This can be written as
\begin{eqnarray}\label{eqn:ad1}
\frac{\overleftrightarrow{n}}{m}=2\sum_K\frac{\partial^2\xi_{\mathbf{k}}}{\partial\mathbf{k}\partial\mathbf{k}}G_K=-2\sum_K\frac{\partial\xi_{\mathbf{k}}}{\partial\mathbf{k}}\frac{\partial G_K}{\partial\mathbf{k}}.
\end{eqnarray}
The right hand side of Eq.~(\ref{eqn:ad1}) can be manipulated so that it appears in a form 
which suggests how to write
$\tensor{P}(Q)$. First differentiating both sides of the equality $G^{-1}_K=G^{-1}_{0,K}-\Sigma_{pg,K}$, one has the identity
\begin{eqnarray}
\frac{\partial G^{-1}_K}{\partial\mathbf{k}}=\frac{\partial G^{-1}_{0,K}}{\partial\mathbf{k}}-\frac{\partial \Sigma_{pg,K}}{\partial\mathbf{k}}=-\frac{\partial\xi_{\mathbf{k}}}{\partial\mathbf{k}}-\frac{\partial\Sigma_{pg,K}}{\partial\mathbf{k}}.
\end{eqnarray}
Using $\partial G_K/\partial\textbf{k}=-G_K^2\partial G_K^{-1}/\partial\bf k$, 
Eq.~(\ref{eqn:ad1}) becomes
\begin{eqnarray}
\frac{\overleftrightarrow{n}}{m}=-2\displaystyle{\sum_K}G^2_K\frac{\partial\xi_{\mathbf{k}}}{\partial\mathbf{k}}\Big[\frac{\partial\xi_{\mathbf{k}}}{\partial\mathbf{k}}+\frac{\partial\Sigma_{pg,K}}{\partial\mathbf{k}}\Big]\label{eq:4b}
\end{eqnarray}
The expression for the self energy, Eq.~(\ref{eq:4a}), can be used to further simplify 
Eq.~(\ref{eq:4b}). Since $\Sigma_{pg,K}=-\Delta_{pg,\bf k}^2G_{0,-K}=\Delta_{pg,\bf k}^2(i\omega_n+\xi_{\mathbf{k}})^{-1}$, then
\begin{eqnarray}
\frac{\partial\Sigma_{pg,K}}{\partial\mathbf{k}}=-\Delta_{pg,\bf k}^2G^2_{0,-K}\frac{\partial\xi_{\mathbf{k}}}{\partial\mathbf{k}},
\end{eqnarray}
where a term proportion to $\partial\Delta_{pg,\textbf{k}}/\partial\textbf{k}$ has been dropped since it gives a negligible contribution to the final result. Therefore Eq.~(\ref{eq:4b}) becomes
\begin{eqnarray}
\frac{\overleftrightarrow{n}}{m}&=&-2\displaystyle{\sum_K}G^2_K\frac{\partial\xi_{\mathbf{k}}}{\partial\mathbf{k}}\frac{\partial\xi_{\mathbf{k}}}{\partial\mathbf{k}}\big(1-\Delta_{pg,\textbf{k}}^2G^2_{0,-K}\big)
\end{eqnarray}
Note that the combination $GG_0$ appears naturally in the manipulations, and
that this same $GG_0$ contribution forms the basis for a t-matrix
ladder summation as summarized in Appendix A.
In order for the Meissner effect to be present only below $T_c$
we require 
\begin{equation}
\tensor{P}(0)+\tensor{n}/m=0,\ T\geq T_c
\label{eq:9}
\end{equation}
which results in
\begin{eqnarray}\tensor{P}(0)&=& {2}\displaystyle{\sum_K}\frac{\partial\xi_{\textbf{k}}}{\partial\textbf{k}}\frac{\partial\xi_{\textbf{k}}}{\partial\textbf{k}}
\Big[G_KG_{K}\big(1  -\Delta_{pg,\textbf{k}}^2G^2_{0,-K}\big) \Big]\label{eq:int}
\end{eqnarray}
A natural extension of Eq.~(\ref{eq:int}) to general $Q$
is
\begin{eqnarray}
{\tensor{P}}(Q)&=& {2}\sum_K\frac{\partial\xi_{\textbf{k}+\textbf{q}/2}}
{\partial\textbf{k}}\frac{\partial\xi_{\textbf{k}+\textbf{q}/2}}
{\partial\textbf{k}}\Big[G_KG_{K+Q}\nonumber\\
&-&\Delta_{pg,\textbf{k}}\Delta_{pg,\textbf{k}+\textbf{q}}G_{0,-K-Q}G_{0,-K}
G_{K+Q}G_K\Big].
\label{eq:int2}
\end{eqnarray}
This ansatz will be checked by appealing to the transverse f-sum rule.
First we rewrite Eq.~(\ref{eq:int2}) as
\begin{widetext}
 \begin{eqnarray}
\label{eq:12}{P}_{xx}(\textbf{q},\omega)=\displaystyle{\sum_{\textbf{k}}}\frac{\partial\xi_{
\text{\bf
k}}}{\partial k_{x}}\frac{\partial\xi_{\text{\bf k}}}{\partial
k_{x}}\Bigg[\frac{E^+_{\textbf{k}}+E^-_{\textbf{k}}}{E^+_{\textbf{k}}E^-_{\textbf{k}}}\frac{E^+_{\textbf{k}}E^-_{\textbf{k}}-\xi^+_{\textbf{k}}\xi^-_{\textbf{k}}-\delta\Delta_{\textbf{k},\textbf{q}}^2}{\omega^2-(E^+_{\textbf{k}}+E^-_{\textbf{k}})^2}\Big(1-f(E^+_{\textbf{k}})-f(E^-_{\textbf{k}})\Big)\\-\frac{E^+_{\textbf{k}}-E^-_{\textbf{k}}
}{E^+_{\textbf{k}}E^-_{\textbf{k}}}\frac{E^+_{\textbf{k}}E^-_{\textbf{k}}+\xi^+_{\textbf{k}}\xi^-_{\textbf{k}}+\delta\Delta_{\textbf{k},\textbf{q}}^2}{\omega^2-(E^+_{\textbf{k}}-E^-_{\textbf{k}}
)^2}\Big(f(E^+_{\textbf{k}})-f(E^-_{\textbf{k}})\Big)\Bigg]\nonumber
\end{eqnarray}
\end{widetext}
where a $\pm$ superscript indicates that the given function is evaluated at $\textbf{k}\pm\textbf{q}/2$. 
We define \begin{eqnarray}\delta\Delta_{\textbf{k},\textbf{q}}^2=-\Delta_{pg,\textbf{k}}^+\Delta_{pg,\textbf{k}}^-,\ T\geq T_c\label{eq:dd}\end{eqnarray} Once the temperature passes below $T_c$, we need to include the self energy of the condensed pairs as well
\begin{eqnarray}
\Sigma_K=\Sigma_{sc,K}+\Sigma_{pg,K}=-\Big[\Delta_{pg,\textbf{k}}^2+\Delta_{sc,\textbf{k}}^2\Big]G_{0,-K}
\end{eqnarray}
where $\Sigma_K$ now consists of a condensed and non-condensed
pair contributions. This results in an expression for the diamagnetic contribution, just as in 
Eq.~(\ref{eq:9}) which
can be rewritten in the form 
\begin{eqnarray}
\label{eq:numdens}\frac{\tensor{n}}{m}=2\displaystyle{\sum_{\textbf{k}}}\frac{\partial\xi_{\mathbf{k}}}{\partial \bf k}\frac{\partial\xi_{\mathbf{k}}}{\partial \bf k}\Big[\frac{\Delta_{\textbf{k}}^2}{E_{\textbf{k}}^2}\frac{1-2f(E_{\textbf{k}})}{2E_{\textbf{k}}}-\frac{\xi_{\textbf{k}}^2}{E_{\textbf{k}}^2}\frac{\partial f(E_{\textbf{k}})}{\partial E_{\textbf{k}}}\Big]
\end{eqnarray}
where $\Delta_{\textbf{k}}^2=\Delta_{pg,\textbf{k}}^2+\Delta_{sc,\textbf{k}}^2$.  
To determine how $\Delta_{sc,\textbf{k}}^2$ enters into the paramagnetic
current $P(Q)$, we observe that, in the BCS limit, $$\delta\Delta_{\textbf{k},\textbf{q}}^2=\Delta_{sc,\textbf{k}}^+\Delta_{sc,\textbf{k}}^-,~~~~ \textrm{BCS limit}$$

An essential point is that the
superconducting gap $\Delta_{sc,\textbf{k}}$ appears with the opposite 
sign from the pseudogap contribution in Eq.~(\ref{eq:dd}). This is
necessary in order to yield a Meissner effect which
disappears when the order parameter disappears.
One can interpret this sign change as associated with
the appropriate vertex corrections.
In the case of general temperatures, $0\leq T\leq T*$, we 
combine the two limits to yield the appropriate form for the quantity 
\begin{eqnarray}\label{eq:dd2}\delta\Delta_{\textbf{k},\textbf{q}}^2=\Delta_{sc,\textbf{k}}^+\Delta_{sc,\textbf{k}}^--\Delta_{pg,\textbf{k}}^+\Delta_{pg,\textbf{k}}^-
\end{eqnarray} 
which enters into Eq.~(\ref{eq:12}). 
Importantly, Eqs.~(\ref{eq:12}) and (\ref{eq:dd2}) represent
the full electromagnetic response above and below $T_c$,
albeit in the weak dissipation limit. 
The superfluid density follows from the definition 
 \begin{eqnarray}\tensor{P}(0)+\tensor{n}/m=\tensor{n}_s/m\label{eq:ns}\end{eqnarray}
Combining Eq.~(\ref{eq:12}),
(\ref{eq:dd2}) 
and (\ref{eq:numdens}) implies that
the superfluid density is given by
\begin{eqnarray}
\frac{n_s}{m}=2\displaystyle{\sum_{\mathbf{k}}}\frac{\partial\xi_{\mathbf{k}}}{\partial k_x}\frac{\partial\xi_{\mathbf{k}}}{\partial k_x}\frac{\Delta^2_{\textrm{sc},\bf k}}{E^2_{\mathbf{k}}}\Big[\frac{1-2f(E_{\mathbf{k}})}{2E_{\mathbf{k}}}+\frac{\partial f(E_{\mathbf{k}})}{\partial E_{\mathbf{k}}}\Big]\label{eq:ns}
\end{eqnarray}
Thus the normal fluid density, which will be used as input into the f-sum rule 
that constrains $\sigma_1(\omega)$, is
$n_n/m = n/m-n_s/m =$
\begin{eqnarray}
2\displaystyle{\sum_{\mathbf{k}}}\frac{\partial\xi_{\mathbf{k}}}{\partial k_x}\frac{\partial\xi_{\mathbf{k}}}{\partial k_x}\Big[\frac{\Delta^2_{\textrm{pg},\textbf{k}}}{E_{\textbf{k}}^2}\frac{1-2f(E_{\mathbf{k}})}{2E_{\mathbf{k}}}-\frac{E^2_{\mathbf{p}}-\Delta^2_{\textrm{pg},\textbf{k}}}{E_{\textbf{k}}^2}\frac{\partial f(E_{\mathbf{k}})}{\partial E_{\mathbf{k}}}\Big]\nonumber
\end{eqnarray}

The transverse f-sum rule is given by
\begin{eqnarray}\label{rulechiT}
\lim_{\textbf{q}\rightarrow0}\int_{-\infty}^{+\infty}\frac{d\omega}{\pi}\big(-\frac{\textrm{Im}
P_{xx}(\textbf{q},i\Omega_m\rightarrow\omega+i0^+) }{\omega}\big)=\frac{n_n}{m},
\end{eqnarray}

This sum rule can be proven to hold analytically by directly using  
Eq.~(\ref{eq:12}), along with the normal fluid density.
From Eq.~(\ref{eq:12}), we have
\begin{widetext}
\begin{eqnarray}
\lim_{q\rightarrow0}\int_{-\infty}^{+\infty}\frac{d\omega}{\pi}\big(-\frac{\textrm{Im}
P_{xx}(\textbf{q},i\Omega_m\rightarrow\omega+i0^+) }{\omega}\big)
&=&\sum_{\mathbf{p}}\frac{1}{2}\frac{\partial\xi_{\mathbf{k}}}{\partial k_x}\frac{\partial\xi_{\mathbf{k}}}{\partial k_x}\Big[\frac{E^2_{\mathbf{k}}-E^2_{\mathbf{k}}+2\Delta^2_{\textrm{pg},\textbf{k}}}{E^2_{\mathbf{k}}}\big(\frac{1}{2E_{\mathbf{k}}}-\frac{1}{-2E_{\mathbf{k}}}\big)\\&\times&\big(1-2f(E_{\mathbf{k}})\big)-2\frac{E^2_{\mathbf{k}}+E^2_{\mathbf{k}}-2\Delta^2_{\textrm{pg},\textbf{k}}}{E^2_{\mathbf{k}}}\lim_{q\rightarrow0}\frac{f(E^+_{\mathbf{k}})-f(E^-_{\mathbf{k}})}{E^+_{\mathbf{k}}-E^-_{\mathbf{k}}}\Big]\nonumber\\
&=&2\displaystyle{\sum_{\mathbf{k}}}\frac{\partial\xi_{\mathbf{k}}}{\partial k_x}\frac{\partial\xi_{\mathbf{k}}}{\partial k_x}\Big[\frac{\Delta^2_{\textrm{pg},\textbf{k}}}{E_{\textbf{k}}^2}\frac{1-2f(E_{\mathbf{k}})}{2E_{\mathbf{k}}}-\frac{E^2_{\mathbf{p}}-\Delta^2_{\textrm{pg},\textbf{k}}}{E_{\textbf{k}}^2}\frac{\partial f(E_{\mathbf{k}})}{\partial E_{\mathbf{k}}}\Big]\nonumber=\frac{n_n}{m}.
\end{eqnarray}
\end{widetext}

Importantly one can see by direct Kramers Kronig analysis that
Eq.~(\ref{eq:9}), which reflects the absence of a normal state
Meissner effect, is intimately connected to the sum rule above $T_c$.

The confirmation of the sum rule then serves to validate Eq.~(\ref{eq:12}),
where importantly
Eq.~({\ref{eq:dd2}) must be used.
We stress that in the usual BCS-like, purely fermionic Hamiltonian (which we
consider here) only fermions
possess a hopping kinetic energy and
thereby directly
contribute to transport, as indicated by the right hand side
of the sum rule. The contribution to transport from
pair correlated fermions
enters
indirectly by liberating these fermions through a break-up of the pairs.

We now see that the general form of the superconducting
electromagnetic response
consists of three distinct contributions:
(1) superfluid acceleration,
(2) quasi-particle
scattering,
and (3) pair breaking and pair forming.
These all appear in conventional BCS superconductors,
but at $T = 0$ this last effect is only present when there
is disorder.
However, in the presence of stronger than BCS attraction
and at $T \neq 0$, non-condensed pairs
can be decomposed to add to the higher frequency conductivity.

\begin{figure*}
\includegraphics[width=5in] {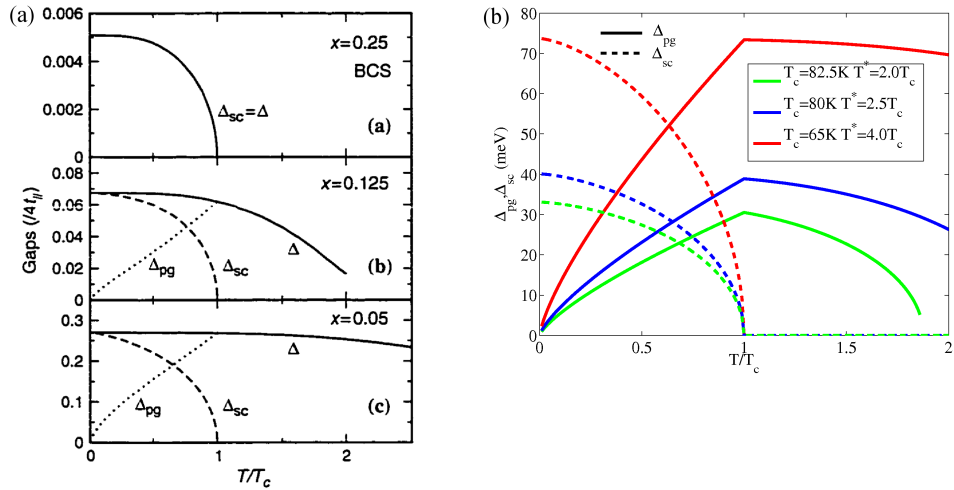}
\caption{(a) The superconducting gap $\Delta_{sc}$ and pseudogap $\Delta_{pg}$ at
the antinode for three different dopings and in units of the in-plane hopping integral $t_
{||}$, obtained self-consistently within
the microscopic model discussed here for a nearest neighbor tight-binding dispersion , as
a function of temperature. Temperature
is measured in units of the transition temperature $T_c$. Solid lines
show $\Delta$, dotted lines $\Delta_{pg}$, and dashed lines $\Delta_{sc}$.
From Ref. \onlinecite{Chenthesis}.
(b) The gaps used for the present calculations.
Superconducting gap $\Delta_{sc}$ and pseudo gap $\Delta_{pg}$ at
the antinode in meV for three different dopings as functions of temperature. Temperature
is measured in units of the transition temperature $T_c$. The solid lines
show $\Delta_{pg}$ and dashed lines denote $\Delta_{sc}$. Here
$\Delta^2 = \Delta_{sc}^2 +\Delta_{pg}^2$ represents the square of
the excitation gap. Details of these parameters are included Ref.\onlinecite{ourarpes}}
\label{fig:sup_2}
\end{figure*}

\subsection{Strong dissipation limit}

We now use the full expression for the self energy
to obtain compatible expressions for transport coefficients in the strong dissipation limit \cite{Arcstransport}. 
The full
Green's function is given by 
\begin{eqnarray} G_K=\Big(i\omega_n-\xi_{\textbf{k}}
+i\gamma-\frac{\Delta_{pg,\textbf{k}}^2}{i\omega_n+\xi_{\textbf{k}}+i\gamma}-
\frac{\Delta_{sc,\textbf{k}}
^2}{i\omega_n+\xi_{\textbf{k}}}\Big)^{-1} 
\end{eqnarray}
Below $T_c$ we introduce
terms
of the form $ F_{sc,K}F_{sc,K+Q}$
which represent the usual Gor'kov functions to represent
the condensate. More specifically,
$F_{sc,K}$
can be represented as a product of one dressed
and one bare Green's function ($GG_0$)
\begin{equation}\label{eq:Fsc}
F_{sc,K} \equiv
 -\frac{\Delta_{sc,\textbf{k}}}{i\omega_n+\xi_{\textbf{k}} } \frac{1}{i\omega_n - \xi_{\textbf{k}}
-\frac{\Delta^2_{\textbf{k}}}{i\omega_n+\xi_{\textbf{k}} }}. 
\end{equation}
This natural extension of our small dissipation result leads to 
\begin{eqnarray}
\tensor{P}(Q)&\approx& {2}\sum_K\frac{\partial\xi_{\textbf{k}+\textbf{q}/2}}{\partial\textbf{k}}\frac{\partial\xi_{\textbf{k}+\textbf{q}/2}}{\partial\textbf{k}}\\ \nonumber
&\times& \left[G_KG_{K+Q} + F_{sc,K}F_{sc,K+Q} -F_{pg,K}F_{pg,K+Q} \right],
\label{eq:int3} 
\end{eqnarray}
where $F_{pg,K} \equiv - \Delta_{pg,\textbf{k}}(i\omega_n+\xi_{\textbf{k}}+ i\gamma)^{-1}G_K$. 
Here the
$F_{pg,K}$ terms represent the non-condensed pair contribution to transport,
which appeared in our small dissipation derivation as well. They
are not to be
associated with broken symmetry. This is, in part,
reflected in the incorporation of the finite lifetime $\gamma^{-1}$ in the expression for $F_{pg,K}$. Rather they represent
correlations among pairs of fermions.
This is in contrast to the $F_{sc,K}$ contributions, which are present only for $T\leq T_c$ and reflect a non-zero order parameter $\Delta_{sc,\textbf{k}}$. 
Note that 
the difference in the relative signs of $\Delta_{pg,\textbf{k}}^2$ and $\Delta_{sc,\textbf{k}}^2$ that appears in Eq.~(\ref{eq:int3}) is a direct consequence of the same physics discussed
in our weak dissipation
calculations.
That the condensed and non-condensed pairs enter in a different fashion is
a crucial finding and one that is essential in order that the pseudogap self
energy does not
contribute to a Meissner effect.

The origin of the fermionic inverse lifetime $\gamma$ 
was discussed very early on \cite{Maly1,Maly2}. In a microscopic t-matrix
theory \cite{ourreview} one considers only pairs (represented by the t-matrix)
and particles (represented by the Green's function $G$) and no higher
order coupling. Then the parameter $\gamma$ arises from the inter-conversion of 
fermions and pairs.

\begin{figure*}
\includegraphics[width=6.25in,clip]
{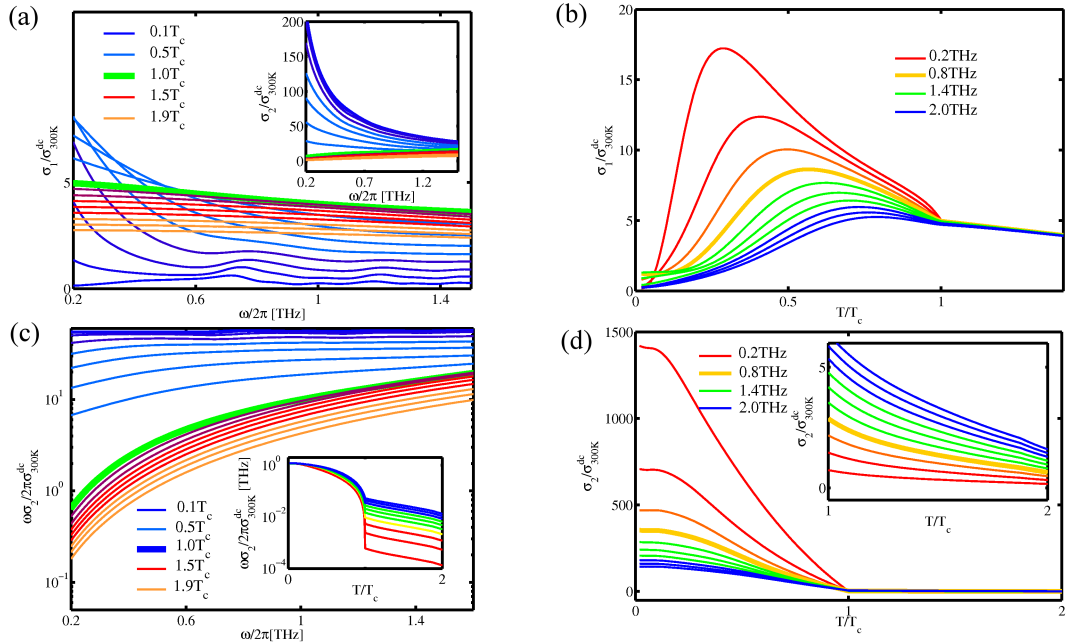}
\caption{(a) The real conductivity $\sigma_1$ as a function of frequency normalized by the dc conductivity at $T=300K$, $\sigma_{300K}^{dc}$. Inset: The imaginary conductivity $\sigma_2$ as a function of frequency.
(b)The real conductivitiy $\sigma_1$ as a function of temperature.
(c) The quantity $\omega\sigma_2$ as a function of frequency. Inset: $\omega\sigma_2$ as a function of temperature.
(d) The imaginary conductivity $\sigma_2$ as a function of temperature. Inset: $\sigma_2$
as a function of temperature near $T_c$.}
\label{fig:2}
\end{figure*}

\section{Calculation of the Pairing Gaps}

Throughout this paper we have implicitly presumed that the gap components
$\Delta_{pg}(T)$ and $\Delta_{sc}(T)$ are known, where the gaps are assumed to be d-wave and $\Delta_{pg}(T)$ and $\Delta_{sc}(T)$ are the gap magnitudes at the antinodes. We now discuss the way
in which these are calculated, referring the reader
to Appendix A for more details.

We consider a preformed pair scenario
which is based
on BCS-Bose Einstein condensation (BCS-BEC) theory.
Given the small pair size and the anomalously high transition
temperatures of the cuprates, one might associate these findings
with a stronger than BCS attractive interaction.
Importantly, the BCS ground state wavefunction
\begin{equation}
|\Psi_0\rangle=\displaystyle{\prod_{\bf k}}(\uk+\vk c_{\textbf{k},\uparrow}^{\dagger} c_{-\textbf{k},\downarrow}^{\dagger})|0\rangle
\label{eq:1a}
\end{equation}
is well known to contain both the BCS and BEC limits. We present in Appendix A
a treatment of finite temperature effects which is based on a t-matrix
implementation of BCS-BEC theory.
Ours represents
a straightforward extension of standard BCS and Gor'kov theory.
Given that we start with the same wavefunction, it is
not surprising that our
pairing scenario is a mean-field scheme just as in strict BCS theory.
Beyond this BCS endpoint there are two types of excitations, fermionic
quasi-particles and pair excitations. The fermions have the usual
dispersion relation
$E_{\bf k}$,
where $E_{\bf k} \equiv \sqrt{\xi_{\textbf{k}} + \Delta^2_{\textbf{k}}}$
and where the excitation gap consists of two contributions
from non-condensed (pg) and condensed (sc) pairs:
via $\Delta^2_{\textbf{k}} \equiv  \Delta_{pg,\textbf{k}}^2 + \Delta_{sc,\textbf{k}}^2$.
We stress that
the preformed pairs represent pair correlations of fermions which
have nothing to do with broken symmetry.
Note that the full gap $\Delta_{\textbf{k}}$ remains relatively T-independent, even below $T_c$ because of the conversion of non-condensed ($\Delta_{pg,\textbf{k}}$) to condensed
($\Delta_{sc,\textbf{k}}$) pairs as the temperature is lowered.

We further amplify the simple physics.
Written in terms of fermion creation and annihilation operators
($c^\dagger$ and $c$ respectively), these pair correlations correspond to
$[\langle cc^\dagger cc^\dagger\rangle
-\langle c^\dagger c^\dagger\rangle
\langle cc\rangle]$ and are
ignored in BCS theory (where the attraction is very weak).
In a closely related fashion, the
(square of the) contribution to the total pairing
gap ($\Delta(T)$)
associated with non-condensed pairs (pg), can be written \cite{CSTL05} as
$\Delta_{pg}^2(T) = [ \Delta^2(T) -\Delta_{sc}^2(T)]$
where sc corresponds to
condensed pairs and pg corresponds to the preformed (pseudogap) pairs.

The results of a full numerical solution\cite{Chenthesis} for these gap
parameters (associated with Eqs.~(\ref{eq:pgeq}), (\ref{eq:BCSgap}), and (\ref{eq:neq}))
for a nearest-neighbor tight-binding dispersion is shown in the  Fig.\ref{fig:sup_2}(a), where the gaps $\Delta$, $\Delta_{sc}$, and $\Delta_{pg}$ are plotted
as functions of temperature and for different dopings, as represented by different interaction
coupling constants. For the calculations performed in this paper, the specific parameters that were used are illustrated in Fig.\ref{fig:sup_2}(b). These particular parameters were chosen for consistency with the cuprate phase diagram,
so that, for example, the attractive interaction was chosen to fit $T^*$. This procedure
is described in more detail in [\onlinecite{ourarpes}].

\section{Detailed Numerical Studies}
We now turn to more detailed comparisons between THz theory and experiment.
Fig.\ref{fig:2} displays our more
quantitative results for $\sigma_1$ and $\sigma_2$ both as functions of
$\omega$ and $T$. 
Our numerical results, based on
Eq.~\eqref{eq:conddef}, are presented
in a layout designed to mirror Figure 1 from Ref. \onlinecite{Bilbro}
where the general trends are similar. 
One sees from Fig.\ref{fig:2}(a) and its inset
that well above $T_c$, the real part of the conductivity is almost
frequency independent. The imaginary part is small in this regime. At the lowest
temperatures $\sigma_1$ contains much reduced spectral weight while the
frequency dependence of $\sigma_2 \propto \omega^{-1}$; both of these reflect the
characteristic behavior
of a superfluid.
The behavior below $T_c$ is not superficially different from
that of strict BCS theory. However, it should be noted that
the pairing gap $\Delta(T)$ (at the antinodes)
is almost $T$ independent. 
BCS theory (which considers only fermionic
excitations) would, thus, predict no significant
T dependences in $\sigma_1$ and $\sigma_2$.

Here, as in the experimental studies \cite{Bilbro},
we focus primarily on the temperature dependent plots
in Figs.\ref{fig:2}(b), (d) and the inset to (c).
One sees that $\sigma_1$ shows a slow decrease as the temperature is
raised above $T_c$.
Somewhat below $T_c$, $\sigma_1$ exhibits a peak that occurs at progressively lower
temperatures as the probe frequency is decreased. At roughly $T_c$, we find
that $\sigma_2$ shows a sharp upturn at low $\omega$. The region of finite $\sigma_2$
above the transition can be seen from the inset in
Fig.\ref{fig:2}(d).
The inset shows an expanded view of
$\sigma_2(T)$ near $T_c$.
In agreement with
experiment, the nesting of the $\sigma_2$ versus T curves switches orders above
$T_c$.

These effects are made clearer by plotting the
``phase stiffness", which is proportional to the quantity
$\omega\sigma_2$
and is shown in Fig.\ref{fig:2}(c). Deep in the superconducting state
there is no $\omega$ dependence to $\omega \sigma_2(\omega)$, while
at higher $T$ this dependence becomes apparent.
In the inset to (c), the temperature
dependence of $\omega\sigma_2(T)$ is displayed.
We see that
above $T_c$,
$\omega \sigma_2$ is never strictly constant, as would
be expected from fluctuation contributions.

In general, these curves capture the qualitative features observed
in recent experiments \cite{Bilbro}.

\section{Conclusions}

In this paper we have shown how the standard self energy expression
($\Sigma_{pg,K}$)  which appears in 
Eq.~(\ref{eq:4a}) and which is widely adopted in the literature
\cite{CSTL05,FermiArcs,NormanTransport}, 
can be used to derive the frequency dependent conductivity $\sigma(\omega)$.
Elsewhere in the literature \cite{ourreview,Kosztin2,Chen2} this
transport approach has
been derived from a more microscopic formalism, (which involves
Maki-Thompson and Aslamazov Larkin diagrams).
Importantly, the results can be seen to be analytically
compatible with the transverse f-sum rule, and semi-quantitatively
compatible with the data. In the normal state
this sum rule constraint is equivalent to the requirement that there is
no Meissner effect. This theory is readily extended below $T_c$
by including a second component to the excitation gap associated with
condensed pairs which is of the usual BCS (undamped) form.
We have additionally shown that the recent experiments by Bilbro et al.\cite{Bilbro}
can be successfully addressed in this framework which
can be microscopically associated with BCS-BEC crossover theory.
Importantly, this particular variant of a preformed pair approach
has
been 
unambiguously realized in (atomic physics) experiments where
a pseudogap is claimed to be observed \cite{Jin6}.

We can summarize the effects of a pseudogap in the
normal state, which differentiates the present theory from that
of its BCS counterpart. In the low frequency regime, with a pseudogap present, there are fewer fermions available to contribute to
transport since their number is reduced because they are
tied up into pairs.
However, once the frequency is sufficiently high
to break the pairs into individual fermions,
the conductivity rises above that of the Drude model.
One can see that the effect of the pseudogap is to transfer the spectral
weight from low frequencies to higher energies, $\omega \approx 2 \Delta$,
(where $\Delta$ is the pairing gap).
In this way one finds an extra ``mid-infrared" contribution to the
conductivity \cite{OurAC} which is as observed \cite{AndoRes1} experimentally
and is strongly tied to the presence of a pseudogap.
This contribution is not, however, visible in the low $\omega$
THz experiments that are considered in later figures.
It is, however, discussed in the following paper.

The behavior of
$\sigma_2(\omega)$ is rather similarly
constrained.
On general principles, $\sigma_2$ must vanish at strictly zero frequency as long
as the system is normal.
Here one can see that the low frequency behavior is also suppressed by the
presence of a pseudogap because of the gap-induced decrease in the number of carriers.
At higher $\omega \approx 2 \Delta$), the second peak in $\sigma_1(\omega)$
leads, via a Kramers-Kronig transform
to a slight depression in $\sigma_2(\omega) $ in this frequency range.
As a result, $\sigma_2(\omega)$ is significally reduced
relative to the Drude result.

We now turn to the question of 
to what extent does the conductivity below
the transition temperature differ from that in strict BCS theory.
Here it is important to stress the complexity of the superfluid 
phase in the presence of a pseudogap. Angle resolved photoemission
experiments
\cite{ShenNature} indicate that the (anti-nodal) 
spectral gap is not sensitive to $T_c$. In strict BCS
theory with a constant pairing gap, the superfluid density
should not vanish at $T_c$. Rather it would vanish when the
excitation gap disappeared, say at $T^*$.
Moreover, since
$\sigma_2 \propto n_s/\omega$
it would then seem to be difficult to understand the behavior
of the THz conductivity which reflects $T_c$ and not $T^*$.

There has to be, therefore, a substantial effect of the
pseudogap which persists below $T_c$, thus differentiating
these systems from conventional BCS superconductors.
In the present theory this difference is incorporated by
including a persistent pseudogap below the transition.
This non-zero $\Delta_{pg}$ is to be associated with 
non-condensed pairs which are present above $T_c$ and
do not immediately disappear once the transition line is
crossed. Rather these non-condensed pairs gradually
convert to the condensate as $T \rightarrow 0$.
As a consequence, in the present approach
we find that
the spectral gap exhibits the $T$-insensitivity
at the anti-nodes \cite{ourarpes} while $n_s$ \cite{Chen2} vanishes
at $T_c$ and appears clearly in
transport.

Finally, we raise the important issue of concomitant order in the 
above $T_c$ pseudogap phase.
Interestingly, we have found such order
to exist in high magnetic fields,
in the form of bosonic charge density wave-like
states or precursor vortex configurations.
Future work will be required to see if this is a more general phenomenon.
Nevertheless,
it should be clear that the THz conductivity and even the two-gap
physics observed in ARPES \cite{ShenNature} are not incompatible
with a preformed pair scenario for the cuprates. They, thus, do not
necessarily require the presence of another order parameter.
\vskip5mm
This work is supported by NSF-MRSEC Grant
0820054. We thank Hao Guo and Chih-Chun Chien, along with
Peter Scherpelz and A. Varlamov for useful
conversations, and L. Bilbro and N. P. Armitage for sharing
unpublished data.

\appendix

\section{Summary of T-matrix Theory}

In this section we summarize previous work \cite{ourreview,Kosztin2,Chen2}
which established a microscopic description of the pseudogap
based on BCS-BEC theory. 
Alternative formulations of preformed pairs of a different
nature from our work are discussed in Ref.\cite{Loktev_Papers}. In the
present paper a
stronger than BCS attraction
leads to boson-like excitations or meta-stable, long lived pairs with
non- zero net momentum.These pairs give rise to a gap for fermionic excitations.  At the microscopic level these
pairs are associated with a t-matrix which is coupled to
the fermionic Green's function, which is, in turn, dependent on
the t-matrix.


It is useful to begin by reformulating strict BCS theory 
as a BEC phenomenon which motivates our extension to treat a stronger than BCS attraction.
Important here
is that
BCS theory can be viewed as
incorporating \textit{virtual} non-condensed pairs.
Here we consider the general case applicable to both
$s$ and $d$-wave pairing by defining the form factor
$\varphi_{{\bf
k}}=[\cos(k_x)-\cos(k_y)]$ for the latter and taking
it to be unity for the former.
%
%
%
%
These virtual $ Q \neq 0$ pairs are associated with
an effective
propagator or t-matrix which is taken to be of the form
\begin{equation}
t(Q) \equiv \frac {U} { 1 + U \sum_{K} G_K G_{0,-K+Q}\varphi_{{\bf
k-q/2}}^{2}}. \label{eq:t}
\end{equation}
in order to yield the standard BCS equations. This t-matrix incorporates
a summation of ladder diagrams in the
particle-particle channel and
importantly depends on
both 
$G$ and $G_0$, which represent dressed and non-interacting Green's
functions respectively. That one has this mixture of
the two Green's functions can be traced back to the gap
equation of Gor'kov theory. In order to describe pairing in the
$d_{x^2-y^2}$-wave channel, we write the attractive fermion-fermion interaction
in the form $U_{{\bf k},{\bf k}^{\prime}}=U\varphi_{{\bf
k}}\varphi_{{\bf k}^{\prime}}$, where $U$ is the strength of the
pairing interaction.
As in bosonic theories, non-condensed pair excitations of the
condensate are necessarily gapless below $T_c$. This means that
$t(Q\rightarrow0)\rightarrow \infty$  and is equivalent to the vanishing of
the effective pair chemical potential $\mu_{pair}$ for $T \leq T_c$. This leads to
a central constraint on the $T$-matrix
$t^{-1}(Q\rightarrow0) = 0 \rightarrow~ \mu_{pair} =0, T \leq T_c $.
In order to identify the above condition with the BCS gap
equation, we need to incorporate the appropriate form for $G_K$. In BCS theory
the fermionic self energy that appears in the fully dressed Green's
function, $G_K$, is 
\begin{eqnarray}
\Sigma_{sc,K} &=& \sum_Q \big[t_{sc}(Q) 
\varphi_{{\bf k}-{\bf q}/2}^{2} \big]
G_{0,-K+Q}
 \\ \nonumber
&=&-\sum_Q \big[\Delta_{sc,{\bf k}}^2 \delta(Q) \big] G_{0,-K+Q} \\ \nonumber
&=& -\Delta_{sc,{\bf k}}^2 G_{0,-K}
\label{eq:Sigsc}
\end{eqnarray}
where $\Delta_{sc,{\bf k}}(T)\equiv\Delta_{sc}(T)\varphi_{{\bf k}}$
is the superconducting order parameter. The full Green's function is
then 
$G_K^{-1}=G_{0,K}^{-1}-\Sigma_{sc,K}$, which, when inserted in
Eq.~(\ref{eq:t}) yields the BCS gap equation below $T_c$
$1 = - U \sum_{\bf k}
\frac{1 - 2 f (E_{\bf k}^{sc})} { 2 E_{\bf k}^{sc}}\varphi_{\bf k}^{2}$
with
$E_{\bf k}^{sc} \equiv \sqrt{\xi_{\bf k}^2 + \Delta_{sc,{\bf k}}^2}$.
We have thus used Eq.~(\ref{eq:t}) 
to derive the standard BCS gap
equation within a t-matrix language. Importantly, this demonstrates that we can
interpret this gap equation as a BEC condition. That is, it is
an extended version of the Thouless criterion of the strict
BCS theory that applies for all $ T \leq T_c$. 

In order to extend the t-matrix theory to include a stronger than BCS attraction we presume that the
$Q \neq 0$ pairs are no longer virtual.
The t-matrix in general possesses two contributions: the ${\bf
q}=0$ contribution that gives rise to the condensed
or superconducting pairs and the ${\bf q} \not =0$ contribution
of Eq.~(\ref{eq:t})
that
describes the correlations associated with the non-condensed pairs.
As a result, the fermionic self-energy also possesses two
contributions that are given by
\begin{widetext}
\begin{equation}
\Sigma_K = \sum_{Q} t(Q) G_{0,-K+Q} \varphi_{{\bf k}-{\bf q}/2}^{2} = \displaystyle{\sum_Q} \Big[t_{sc}(Q) + t_{pg}(Q) \Big]
G_{0,-K+Q} \varphi_{{\bf k}-{\bf q}/2}^{2} = \Sigma_{sc,K} + \Sigma_{pg,K}
\label{eq:Sigtotal}
\end{equation}
\end{widetext}
The resulting full Green's function is 
$G^{-1}_K=G_{0,K}^{-1}-\Sigma_{sc,K}-\Sigma_{pg,K}$.
While, as before, $\Sigma_{sc,K} =-\Delta_{sc,{\bf k}}^2 G_{0,-K}$, we find numerically
\cite{Maly1,Maly2} that $\Sigma_{pg,K}$ is in general of the form
\begin{equation}
\Sigma_{pg,K} \approx \frac{\Delta_{pg,{\bf
k}}^2}{\omega+\xi_{\bf k}+i\gamma} 
\end{equation}
with $\Delta_{pg,{\bf k}} = \Delta_{pg} \varphi_{\bf k}$. That is,
the self-energy associated with the non-condensed pairs possesses
the same structure as its BCS counterparts, albeit with
a finite lifetime, $\gamma^{-1}$.

We can understand these results more physically as arising from 
the fact that $t_{pg}(Q)$ is strongly peaked around $Q =0$
below $T_c$ where the pair chemical potential is zero and for
a range of temperatures above $T_c$ as well where this chemical
potential is small. 
Thus
the bulk of the contribution to $\Sigma_{pg,K}$ in the ordered state
comes from small $Q$
\begin{equation}
\Sigma_{pg,K} \approx - 
G_{0,-K}
\sum_Q t_{pg}(Q)
 \label{eq:approxSig}
\end{equation}
If we define
\begin{equation}\label{eq:pgeq}
\Delta_{pg,\textbf{k}}^2 \equiv -\sum_Q t_{pg}(Q)\varphi_{\textbf{k}}^2
\end{equation}
we may write

\begin{equation}
\Sigma_K \approx - (\Delta_{sc,{\bf k}}^2 + \Delta_{pg,{\bf k}}^2)
G_{0,-K} \equiv - \Delta_{\bf k}^2 G_{0,-K} \label{eq:approxSig}
\end{equation}
 
Eq.\ref{eq:approxSig} leads to an effective pairing gap
$\Delta(T)$ whose square is associated with the sum of the squares
of the condensed and non-condensed contributions
\begin{equation}
\Delta_{\bf k}^2(T)=\Delta_{sc,{\bf k}}^2(T) + \Delta_{pg,{\bf k}}^2(T)
\end{equation}
Note that the full gap $\Delta_{\textbf{k}}$ remains relatively T-independent, even below $T_c$ because of the conversion of non-condensed ($\Delta_{pg,\textbf{k}}$) to condensed
($\Delta_{sc,\textbf{k}}$) pairs as the temperature is lowered.
The gap equation for this pairing gap,
$\Delta_{\textbf{k}}(T)=\Delta(T)\varphi_{\textbf{k}}$, is again
obtained from the condition $t_{pg}^{-1}(Q = 0)=0$, and given by
\begin{equation}\label{eq:BCSgap}
1 = - U \sum \frac{1 - 2 f(E_{\bf k})}{2 E_{\bf k}} \varphi_{{\bf
k}}^{2},
\end{equation}
where $E_{\mathbf{k}} \equiv \sqrt{ \xi_{\mathbf{k}}^2 + \Delta^2(T)\varphi_{\textbf{k}}^2  }$ ,and
$f$ is the Fermi distribution function. Note that one needs to
self-consistently determine the fermionic chemical potential, $\mu$, by
conserving the number of particles, $n = 2 \sum_K G_K$, which
leads to
\begin{equation}
n = 2 \sum_KG_K = \sum _{\bf k} \left[ 1 -\frac{\xi_{\bf
k}}{E_{\bf k}} +2\frac{\xi_{\bf k}}{E_{\bf k}}f(E_{\bf k})  \right]
\label{eq:neq}
\end{equation}
Eqs.~(\ref{eq:pgeq}), (\ref{eq:BCSgap}), and (\ref{eq:neq}) present a closed set of
equations for the chemical potential $\mu$, the pairing gap
$\Delta_{\textbf{k}}(T)=\Delta(T)\varphi_{\textbf{k}}$, the pseudogap
$\Delta_{pg,{\bf k}}(T)\equiv\Delta_{pg}(T)\varphi_{{\bf k}}$, and 
the superconducting order
parameter $\Delta_{sc,\textbf{k}}(T)=\Delta_{sc}\varphi_{\textbf{k}}$
with $\Delta_{sc}(T)=\sqrt{\Delta^2(T)-\Delta_{pg}^2(T)}$.  
We find that $\Delta_{pg}(T)$ essentially vanishes at $T=0$ where $\Delta = \Delta_{sc}$. 
In this way, the "two
gap" physics disappears in the ground state. 
Importantly,
numerical studies 
\cite{Chienlattice2} show that 
for $d$-wave pairing, there is no superfluid phase in the
bosonic regime where $\mu$ is negative; the pseudogap
is, thus, associated with the
fermionic regime.
With this as a starting point, transport properties
can then be derived. At the diagrammic level the calculation
involves both the Maki-Thompson and Aslamazov-Larkin diagrams \cite{ourreview,Kosztin2}.

This work is supported by NSF-MRSEC Grant
0820054 and we thank L. S. Bilbro, P. Armitage and A. Varlamov for useful
conversations.

\bibliographystyle{apsrev}
\bibliography{Literature}

\begin{thebibliography}{28}
\expandafter\ifx\csname natexlab\endcsname\relax\def\natexlab#1{#1}\fi
\expandafter\ifx\csname bibnamefont\endcsname\relax
  \def\bibnamefont#1{#1}\fi
\expandafter\ifx\csname bibfnamefont\endcsname\relax
  \def\bibfnamefont#1{#1}\fi
\expandafter\ifx\csname citenamefont\endcsname\relax
  \def\citenamefont#1{#1}\fi
\expandafter\ifx\csname url\endcsname\relax
  \def\url#1{\texttt{#1}}\fi
\expandafter\ifx\csname urlprefix\endcsname\relax\def\urlprefix{URL }\fi
\providecommand{\bibinfo}[2]{#2}
\providecommand{\eprint}[2][]{\url{#2}}

\bibitem[{\citenamefont{Daou et~al.}(2010)\citenamefont{Daou, Chang, LeBoeuf,
  Cyr-Choiniere, Laliberte, Doiron-Leyraud, Ramshaw, Liang, Bonn, Hardy
  et~al.}}]{Taillefer4}
\bibinfo{author}{\bibfnamefont{R.}~\bibnamefont{Daou}},
  \bibinfo{author}{\bibfnamefont{J.}~\bibnamefont{Chang}},
  \bibinfo{author}{\bibfnamefont{D.}~\bibnamefont{LeBoeuf}},
  \bibinfo{author}{\bibfnamefont{O.}~\bibnamefont{Cyr-Choiniere}},
  \bibinfo{author}{\bibfnamefont{F.}~\bibnamefont{Laliberte}},
  \bibinfo{author}{\bibfnamefont{N.}~\bibnamefont{Doiron-Leyraud}},
  \bibinfo{author}{\bibfnamefont{B.~J.} \bibnamefont{Ramshaw}},
  \bibinfo{author}{\bibfnamefont{R.}~\bibnamefont{Liang}},
  \bibinfo{author}{\bibfnamefont{D.~A.} \bibnamefont{Bonn}},
  \bibinfo{author}{\bibfnamefont{W.}~\bibnamefont{Hardy}},
  \bibnamefont{et~al.}, \bibinfo{journal}{Nature}
  \textbf{\bibinfo{volume}{463}}, \bibinfo{pages}{519} (\bibinfo{year}{2010}).

\bibitem[{\citenamefont{{Hinkov} et~al.}(2007)\citenamefont{{Hinkov},
  {Bourges}, {Pailhes}, {Sidis}, {Ivanov}, {Frost}, {Perring}, {Lin}, {Chen},
  and {Keimer}}}]{Hinkov}
\bibinfo{author}{\bibfnamefont{V.}~\bibnamefont{{Hinkov}}},
  \bibinfo{author}{\bibfnamefont{P.}~\bibnamefont{{Bourges}}},
  \bibinfo{author}{\bibfnamefont{S.}~\bibnamefont{{Pailhes}}},
  \bibinfo{author}{\bibfnamefont{Y.}~\bibnamefont{{Sidis}}},
  \bibinfo{author}{\bibfnamefont{A.}~\bibnamefont{{Ivanov}}},
  \bibinfo{author}{\bibfnamefont{C.~D.} \bibnamefont{{Frost}}},
  \bibinfo{author}{\bibfnamefont{T.~G.} \bibnamefont{{Perring}}},
  \bibinfo{author}{\bibfnamefont{C.~T.} \bibnamefont{{Lin}}},
  \bibinfo{author}{\bibfnamefont{D.~P.} \bibnamefont{{Chen}}},
  \bibnamefont{and} \bibinfo{author}{\bibfnamefont{B.}~\bibnamefont{{Keimer}}},
  \bibinfo{journal}{Nature Physics} \textbf{\bibinfo{volume}{3}},
  \bibinfo{pages}{780} (\bibinfo{year}{2007}).

\bibitem[{\citenamefont{Corson et~al.}(1999)\citenamefont{Corson, Mallozzi,
  Orenstein, Eckstein, and Bozovic}}]{Corson1999}
\bibinfo{author}{\bibfnamefont{J.}~\bibnamefont{Corson}},
  \bibinfo{author}{\bibfnamefont{R.}~\bibnamefont{Mallozzi}},
  \bibinfo{author}{\bibfnamefont{J.}~\bibnamefont{Orenstein}},
  \bibinfo{author}{\bibfnamefont{J.~N.} \bibnamefont{Eckstein}},
  \bibnamefont{and} \bibinfo{author}{\bibfnamefont{I.}~\bibnamefont{Bozovic}},
  \bibinfo{journal}{Nature} \textbf{\bibinfo{volume}{398}},
  \bibinfo{pages}{221} (\bibinfo{year}{1999}).

\bibitem[{\citenamefont{Li et~al.}(2010)\citenamefont{Li, Wang, Komiya, Ono,
  Ando, Gu, and Ong}}]{Ong2}
\bibinfo{author}{\bibfnamefont{L.}~\bibnamefont{Li}},
  \bibinfo{author}{\bibfnamefont{Y.}~\bibnamefont{Wang}},
  \bibinfo{author}{\bibfnamefont{S.}~\bibnamefont{Komiya}},
  \bibinfo{author}{\bibfnamefont{S.}~\bibnamefont{Ono}},
  \bibinfo{author}{\bibfnamefont{Y.}~\bibnamefont{Ando}},
  \bibinfo{author}{\bibfnamefont{G.~D.} \bibnamefont{Gu}}, \bibnamefont{and}
  \bibinfo{author}{\bibfnamefont{N.~P.} \bibnamefont{Ong}},
  \bibinfo{journal}{Phys. Rev. B} \textbf{\bibinfo{volume}{81}},
  \bibinfo{pages}{054510} (\bibinfo{year}{2010}).

\bibitem[{\citenamefont{Bilbro et~al.}(2011{\natexlab{a}})\citenamefont{Bilbro,
  Guilar, Logvenov, Pelleg, Bozovic, and Armitage}}]{Bilbro}
\bibinfo{author}{\bibfnamefont{L.~S.} \bibnamefont{Bilbro}},
  \bibinfo{author}{\bibfnamefont{R.~V.} \bibnamefont{Guilar}},
  \bibinfo{author}{\bibfnamefont{B.}~\bibnamefont{Logvenov}},
  \bibinfo{author}{\bibfnamefont{O.}~\bibnamefont{Pelleg}},
  \bibinfo{author}{\bibfnamefont{I.}~\bibnamefont{Bozovic}}, \bibnamefont{and}
  \bibinfo{author}{\bibfnamefont{N.~P.} \bibnamefont{Armitage}},
  \bibinfo{journal}{Nature Physics} \textbf{\bibinfo{volume}{7}},
  \bibinfo{pages}{2980302} (\bibinfo{year}{2011}{\natexlab{a}}).

\bibitem[{\citenamefont{Bilbro et~al.}(2011{\natexlab{b}})\citenamefont{Bilbro,
  ValdesAguilar, Logvenov, Bozovic, and Armitage}}]{Bilbro2}
\bibinfo{author}{\bibfnamefont{L.~S.} \bibnamefont{Bilbro}},
  \bibinfo{author}{\bibfnamefont{R.}~\bibnamefont{ValdesAguilar}},
  \bibinfo{author}{\bibfnamefont{G.}~\bibnamefont{Logvenov}},
  \bibinfo{author}{\bibfnamefont{I.}~\bibnamefont{Bozovic}}, \bibnamefont{and}
  \bibinfo{author}{\bibfnamefont{N.~P.} \bibnamefont{Armitage}},
  \bibinfo{journal}{Phys. Rev. B} \textbf{\bibinfo{volume}{84}},
  \bibinfo{pages}{100511(R)} (\bibinfo{year}{2011}{\natexlab{b}}).

\bibitem[{\citenamefont{Chen et~al.}(2005{\natexlab{a}})\citenamefont{Chen,
  Stajic, Tan, and Levin}}]{CSTL05}
\bibinfo{author}{\bibfnamefont{Q.~J.} \bibnamefont{Chen}},
  \bibinfo{author}{\bibfnamefont{J.}~\bibnamefont{Stajic}},
  \bibinfo{author}{\bibfnamefont{S.}~\bibnamefont{Tan}}, \bibnamefont{and}
  \bibinfo{author}{\bibfnamefont{K.}~\bibnamefont{Levin}},
  \bibinfo{journal}{Phys. Rep.} \textbf{\bibinfo{volume}{412}},
  \bibinfo{pages}{1} (\bibinfo{year}{2005}{\natexlab{a}}).

\bibitem[{\citenamefont{Chen et~al.}(1998)\citenamefont{Chen, Kosztin, Jank\'o,
  and Levin}}]{Chen2}
\bibinfo{author}{\bibfnamefont{Q.~J.} \bibnamefont{Chen}},
  \bibinfo{author}{\bibfnamefont{I.}~\bibnamefont{Kosztin}},
  \bibinfo{author}{\bibfnamefont{B.}~\bibnamefont{Jank\'o}}, \bibnamefont{and}
  \bibinfo{author}{\bibfnamefont{K.}~\bibnamefont{Levin}},
  \bibinfo{journal}{Phys. Rev. Lett.} \textbf{\bibinfo{volume}{81}},
  \bibinfo{pages}{4708} (\bibinfo{year}{1998}).

\bibitem[{\citenamefont{Wulin et~al.}(2011)\citenamefont{Wulin, Fregoso, Guo,
  Chien, and Levin}}]{Arcstransport}
\bibinfo{author}{\bibfnamefont{D.}~\bibnamefont{Wulin}},
  \bibinfo{author}{\bibfnamefont{B.~M.} \bibnamefont{Fregoso}},
  \bibinfo{author}{\bibfnamefont{H.}~\bibnamefont{Guo}},
  \bibinfo{author}{\bibfnamefont{C.-C.} \bibnamefont{Chien}}, \bibnamefont{and}
  \bibinfo{author}{\bibfnamefont{K.}~\bibnamefont{Levin}},
  \bibinfo{journal}{Phys. Rev. B} \textbf{\bibinfo{volume}{84}},
  \bibinfo{pages}{140509(R)} (\bibinfo{year}{2011}).

\bibitem[{\citenamefont{Guo et~al.}(2011{\natexlab{a}})\citenamefont{Guo,
  Wulin, Chien, and Levin}}]{NJOP}
\bibinfo{author}{\bibfnamefont{H.}~\bibnamefont{Guo}},
  \bibinfo{author}{\bibfnamefont{D.}~\bibnamefont{Wulin}},
  \bibinfo{author}{\bibfnamefont{C.-C.} \bibnamefont{Chien}}, \bibnamefont{and}
  \bibinfo{author}{\bibfnamefont{K.}~\bibnamefont{Levin}},
  \bibinfo{journal}{New Journal of Physics} \textbf{\bibinfo{volume}{13}},
  \bibinfo{pages}{075011} (\bibinfo{year}{2011}{\natexlab{a}}).

\bibitem[{\citenamefont{Guo et~al.}(2011{\natexlab{b}})\citenamefont{Guo,
  Wulin, Chien, and Levin}}]{Ourviscosity}
\bibinfo{author}{\bibfnamefont{H.}~\bibnamefont{Guo}},
  \bibinfo{author}{\bibfnamefont{D.}~\bibnamefont{Wulin}},
  \bibinfo{author}{\bibfnamefont{C.-C.} \bibnamefont{Chien}}, \bibnamefont{and}
  \bibinfo{author}{\bibfnamefont{K.}~\bibnamefont{Levin}},
  \bibinfo{journal}{Phys. Rev. Lett.} \textbf{\bibinfo{volume}{107}},
  \bibinfo{pages}{020403} (\bibinfo{year}{2011}{\natexlab{b}}).

\bibitem[{\citenamefont{Guo et~al.}(2010)\citenamefont{Guo, Chien, and
  Levin}}]{OurBraggPRL}
\bibinfo{author}{\bibfnamefont{H.}~\bibnamefont{Guo}},
  \bibinfo{author}{\bibfnamefont{C.-C.} \bibnamefont{Chien}}, \bibnamefont{and}
  \bibinfo{author}{\bibfnamefont{K.}~\bibnamefont{Levin}},
  \bibinfo{journal}{Phys. Rev. Lett.} \textbf{\bibinfo{volume}{105}},
  \bibinfo{pages}{120401} (\bibinfo{year}{2010}).

\bibitem[{\citenamefont{Chen and Levin}(2008)}]{FermiArcs}
\bibinfo{author}{\bibfnamefont{Q.}~\bibnamefont{Chen}} \bibnamefont{and}
  \bibinfo{author}{\bibfnamefont{K.}~\bibnamefont{Levin}},
  \bibinfo{journal}{Phys. Rev. B.} \textbf{\bibinfo{volume}{78}},
  \bibinfo{pages}{020513(R)} (\bibinfo{year}{2008}).

\bibitem[{\citenamefont{Levchenko et~al.}(2010)\citenamefont{Levchenko,
  Micklitz, Norman, and Paul}}]{NormanTransport}
\bibinfo{author}{\bibfnamefont{A.}~\bibnamefont{Levchenko}},
  \bibinfo{author}{\bibfnamefont{T.}~\bibnamefont{Micklitz}},
  \bibinfo{author}{\bibfnamefont{M.~R.} \bibnamefont{Norman}},
  \bibnamefont{and} \bibinfo{author}{\bibfnamefont{I.}~\bibnamefont{Paul}},
  \bibinfo{journal}{Phys. Rev. B} \textbf{\bibinfo{volume}{82}},
  \bibinfo{pages}{060502(R)} (\bibinfo{year}{2010}).

\bibitem[{\citenamefont{Senthil and Lee}(2009)}]{SenthilLee}
\bibinfo{author}{\bibfnamefont{T.}~\bibnamefont{Senthil}} \bibnamefont{and}
  \bibinfo{author}{\bibfnamefont{P.~A.} \bibnamefont{Lee}},
  \bibinfo{journal}{Phys. Rev. B} \textbf{\bibinfo{volume}{79}},
  \bibinfo{pages}{245116} (\bibinfo{year}{2009}).

\bibitem[{\citenamefont{Chubukov et~al.}(2007)\citenamefont{Chubukov, Norman,
  Millis, and Abrahams}}]{Chubukov2}
\bibinfo{author}{\bibfnamefont{A.~V.} \bibnamefont{Chubukov}},
  \bibinfo{author}{\bibfnamefont{M.~R.} \bibnamefont{Norman}},
  \bibinfo{author}{\bibfnamefont{A.~J.} \bibnamefont{Millis}},
  \bibnamefont{and} \bibinfo{author}{\bibfnamefont{E.}~\bibnamefont{Abrahams}},
  \bibinfo{journal}{\prb} \textbf{\bibinfo{volume}{76}},
  \bibinfo{pages}{180501(R)} (\bibinfo{year}{2007}).

\bibitem[{\citenamefont{Maly et~al.}(1999{\natexlab{a}})\citenamefont{Maly,
  Jank\'o, and Levin}}]{Maly1}
\bibinfo{author}{\bibfnamefont{J.}~\bibnamefont{Maly}},
  \bibinfo{author}{\bibfnamefont{B.}~\bibnamefont{Jank\'o}}, \bibnamefont{and}
  \bibinfo{author}{\bibfnamefont{K.}~\bibnamefont{Levin}},
  \bibinfo{journal}{Physica C} \textbf{\bibinfo{volume}{321}},
  \bibinfo{pages}{113} (\bibinfo{year}{1999}{\natexlab{a}}).

\bibitem[{\citenamefont{Maly et~al.}(1999{\natexlab{b}})\citenamefont{Maly,
  Jank\'o, and Levin}}]{Maly2}
\bibinfo{author}{\bibfnamefont{J.}~\bibnamefont{Maly}},
  \bibinfo{author}{\bibfnamefont{B.}~\bibnamefont{Jank\'o}}, \bibnamefont{and}
  \bibinfo{author}{\bibfnamefont{K.}~\bibnamefont{Levin}},
  \bibinfo{journal}{Phys. Rev. B} \textbf{\bibinfo{volume}{59}},
  \bibinfo{pages}{1354} (\bibinfo{year}{1999}{\natexlab{b}}).

\bibitem[{\citenamefont{Kosztin et~al.}(2000)\citenamefont{Kosztin, Chen, Kao,
  and Levin}}]{Kosztin2}
\bibinfo{author}{\bibfnamefont{I.}~\bibnamefont{Kosztin}},
  \bibinfo{author}{\bibfnamefont{Q.~J.} \bibnamefont{Chen}},
  \bibinfo{author}{\bibfnamefont{Y.-J.} \bibnamefont{Kao}}, \bibnamefont{and}
  \bibinfo{author}{\bibfnamefont{K.}~\bibnamefont{Levin}},
  \bibinfo{journal}{Phys. Rev. B} \textbf{\bibinfo{volume}{61}},
  \bibinfo{pages}{11662} (\bibinfo{year}{2000}).

\bibitem[{\citenamefont{Chen et~al.}(2005{\natexlab{b}})\citenamefont{Chen,
  Stajic, Tan, and Levin}}]{ourreview}
\bibinfo{author}{\bibfnamefont{Q.~J.} \bibnamefont{Chen}},
  \bibinfo{author}{\bibfnamefont{J.}~\bibnamefont{Stajic}},
  \bibinfo{author}{\bibfnamefont{S.~N.} \bibnamefont{Tan}}, \bibnamefont{and}
  \bibinfo{author}{\bibfnamefont{K.}~\bibnamefont{Levin}},
  \bibinfo{journal}{Phys. Rep.} \textbf{\bibinfo{volume}{412}},
  \bibinfo{pages}{1} (\bibinfo{year}{2005}{\natexlab{b}}).

\bibitem[{\citenamefont{Chen}(2000)}]{Chenthesis}
\bibinfo{author}{\bibfnamefont{Q.}~\bibnamefont{Chen}}, Ph.D. thesis,
  \bibinfo{school}{University of Chicago} (\bibinfo{year}{2000}).

\bibitem[{\citenamefont{Chien et~al.}(2009)\citenamefont{Chien, He, Chen, and
  Levin}}]{ourarpes}
\bibinfo{author}{\bibfnamefont{C.-C.} \bibnamefont{Chien}},
  \bibinfo{author}{\bibfnamefont{Y.}~\bibnamefont{He}},
  \bibinfo{author}{\bibfnamefont{Q.}~\bibnamefont{Chen}}, \bibnamefont{and}
  \bibinfo{author}{\bibfnamefont{K.}~\bibnamefont{Levin}},
  \bibinfo{journal}{Phys. Rev. B} \textbf{\bibinfo{volume}{79}},
  \bibinfo{pages}{214527} (\bibinfo{year}{2009}).

\bibitem[{\citenamefont{Stewart et~al.}(2008)\citenamefont{Stewart, Gaebler,
  and Jin}}]{Jin6}
\bibinfo{author}{\bibfnamefont{J.~T.} \bibnamefont{Stewart}},
  \bibinfo{author}{\bibfnamefont{J.~P.} \bibnamefont{Gaebler}},
  \bibnamefont{and} \bibinfo{author}{\bibfnamefont{D.~S.} \bibnamefont{Jin}},
  \bibinfo{journal}{Nature} \textbf{\bibinfo{volume}{454}},
  \bibinfo{pages}{744} (\bibinfo{year}{2008}).

\bibitem[{\citenamefont{Wulin et~al.}()\citenamefont{Wulin, Guo, Chien, and
  Levin}}]{OurAC}
\bibinfo{author}{\bibfnamefont{D.}~\bibnamefont{Wulin}},
  \bibinfo{author}{\bibfnamefont{H.}~\bibnamefont{Guo}},
  \bibinfo{author}{\bibfnamefont{C.-C.} \bibnamefont{Chien}}, \bibnamefont{and}
  \bibinfo{author}{\bibfnamefont{K.}~\bibnamefont{Levin}},
  \bibinfo{note}{eprint, arXiv:1108.4375}.

\bibitem[{\citenamefont{Lee et~al.}(2005)\citenamefont{Lee, Segawa, Li,
  Padilla, Dumm, Dordevic, Homes, Ando, and Basov}}]{AndoRes1}
\bibinfo{author}{\bibfnamefont{Y.~S.} \bibnamefont{Lee}},
  \bibinfo{author}{\bibfnamefont{K.}~\bibnamefont{Segawa}},
  \bibinfo{author}{\bibfnamefont{Z.~Q.} \bibnamefont{Li}},
  \bibinfo{author}{\bibfnamefont{W.~J.} \bibnamefont{Padilla}},
  \bibinfo{author}{\bibfnamefont{M.}~\bibnamefont{Dumm}},
  \bibinfo{author}{\bibfnamefont{S.~V.} \bibnamefont{Dordevic}},
  \bibinfo{author}{\bibfnamefont{C.~C.} \bibnamefont{Homes}},
  \bibinfo{author}{\bibfnamefont{Y.}~\bibnamefont{Ando}}, \bibnamefont{and}
  \bibinfo{author}{\bibfnamefont{D.~N.} \bibnamefont{Basov}},
  \bibinfo{journal}{Phys. Rev. B.} \textbf{\bibinfo{volume}{72}},
  \bibinfo{pages}{054529} (\bibinfo{year}{2005}).

\bibitem[{\citenamefont{Lee et~al.}(2007)\citenamefont{Lee, Vishik, Tanaka, Lu,
  Sasagawa, Nagaosa, Devereaux, Hussain, and Shen}}]{ShenNature}
\bibinfo{author}{\bibfnamefont{W.~S.} \bibnamefont{Lee}},
  \bibinfo{author}{\bibfnamefont{I.~M.} \bibnamefont{Vishik}},
  \bibinfo{author}{\bibfnamefont{K.}~\bibnamefont{Tanaka}},
  \bibinfo{author}{\bibfnamefont{D.~H.} \bibnamefont{Lu}},
  \bibinfo{author}{\bibfnamefont{T.}~\bibnamefont{Sasagawa}},
  \bibinfo{author}{\bibfnamefont{N.}~\bibnamefont{Nagaosa}},
  \bibinfo{author}{\bibfnamefont{T.~P.} \bibnamefont{Devereaux}},
  \bibinfo{author}{\bibfnamefont{Z.}~\bibnamefont{Hussain}}, \bibnamefont{and}
  \bibinfo{author}{\bibfnamefont{Z.~X.} \bibnamefont{Shen}},
  \bibinfo{journal}{Nature} \textbf{\bibinfo{volume}{450}}, \bibinfo{pages}{81}
  (\bibinfo{year}{2007}).

\bibitem[{Lok()}]{Loktev_Papers}
\bibinfo{note}{V.M. Loktev, R.M. Quick, and S.G. Sharapov, Phys. Rep.
  \textbf{349}, 1 (2001); V.M. Loktev and S.G. Sharapov, Low Temp. Phys.
  \textbf{23}, 132 (1997); V.M. Loktev, Y.G. Pogorelov, and V.M. Turkowski,
  Int. J. Mod. Phys. B \textbf{17}, 3607 (2003); V.M. Loktev and V.M.
  Turkowski, Low Temp. Phys. \textbf{32}, 802 (2006)}.

\bibitem[{\citenamefont{Chien et~al.}(2008)\citenamefont{Chien, Chen, and
  Levin}}]{Chienlattice2}
\bibinfo{author}{\bibfnamefont{C.~C.} \bibnamefont{Chien}},
  \bibinfo{author}{\bibfnamefont{Q.~J.} \bibnamefont{Chen}}, \bibnamefont{and}
  \bibinfo{author}{\bibfnamefont{K.}~\bibnamefont{Levin}},
  \bibinfo{journal}{Phys. Rev. A} \textbf{\bibinfo{volume}{78}},
  \bibinfo{pages}{043612} (\bibinfo{year}{2008}).

\end{thebibliography}

\end{document}